\documentclass[conference, ]{IEEEtran}
\IEEEoverridecommandlockouts
\usepackage{cite}
\usepackage{amsmath,amssymb,amsfonts}
\usepackage{algorithmic}
\usepackage{graphicx}
\usepackage{textcomp}
\usepackage{xcolor}
\usepackage{hyperref}
\def\BibTeX{{\rm B\kern-.05em{\sc i\kern-.025em b}\kern-.08em
    T\kern-.1667em\lower.7ex\hbox{E}\kern-.125emX}}

\makeatletter
\def\endthebibliography{%
	\def\@noitemerr{\@latex@warning{Empty `thebibliography' environment}}%
	\endlist
}
\makeatother

\begin{document}

\title{Blind Deblurring using Deep Learning: A Survey
\thanks{*denotes equal contribution}}

\author{\IEEEauthorblockN{Siddhant Sahu \textsuperscript{*}}
\IEEEauthorblockA{\textit{dept.Computer Science and Engineering} \\
\textit{KIIT University}\\
Bhubansewar, India \\
hello@siddhantsahu.com}
\and
\IEEEauthorblockN{Manoj Kumar Lenka \textsuperscript{*}}
\IEEEauthorblockA{\textit{dept.Computer Science and Engineering} \\
\textit{KIIT University}\\
Bhubansewar, India \\
manojlenka1998@gmail.com}
\and
\IEEEauthorblockN{Pankaj Kumar Sa}
\IEEEauthorblockA{\textit{dept.Computer Science and Engineering} \\
\textit{NIT Rourkela}\\
Rourkela, India \\
pankajksa@nitrkl.ac.in}
}

\maketitle

\begin{abstract}
We inspect all the deep learning based solutions and provide holistic understanding of various architectures that have evolved over the past few years to solve blind deblurring. The introductory work used deep learning to estimate some features of the blur kernel and then moved onto predicting the blur kernel entirely, which converts the problem into non-blind deblurring. The recent state of the art techniques are end to end i.e they don’t estimate the blur kernel rather try to estimate the latent sharp image directly from the blurred image. The benchmarking PSNR and SSIM values on standard datasets of GOPRO and K\"{o}hler using various architectures are also provided.
\end{abstract}


\section{Introduction} 
Present day imaging systems for instance consumer level photography cameras, medical imaging equipments, scientific astronomical imaging systems, microscopy and more may experience blurring due to various intrinsic (diffraction, lens chromatic aberration, anti-aliasing filters etc.) or extrinsic (object motion, camera shake, out of focus, atmospheric turbulence etc.) factors which results in loss of image information. To overcome this problem and to recover lost information, deblurring is of great interest. From an artistic perspective blur is sometimes intentional in photography but for majority of the image analysis applications blurs ruins useful data. 

The problem of deblurring is restoring a latent sharp image from a blurred image alone or at times with some statistical information about the blur kernel. This has attracted many researchers who have given many different solutions. These solutions can be broadly divided into statistical methods like,
\begin{enumerate}
	\item Bayesian Inference Framework 
	\item Variational Methods 
	\item Sparse Representation based method 
	\item Homography based modeling 
	\item Region based methods 
\end{enumerate}
where we try to estimate the blur kernel from just a single given blurred image and learning based methods (\cite{Chakrabarti}, \cite{Gong_2017_CVPR}, \cite{Kupyn_2018_CVPR}, \cite{Nah_2017_CVPR}, \cite{Noroozi}, \cite{Ramakrishnan}, \cite{Schuler}, \cite{Sun_2015_CVPR}, \cite{Tao_2018_CVPR}, \cite{Zhang_2018_CVPR}) which is data driven and the blur kernel is learned by providing not just one but several examples of blur and its corresponding sharp images as ground truth.

A blurred image can be modeled using equation,
\begin{equation}
	B = K * I + N
\end{equation}
where $B$ is the blurred image, $K$ is the kernel, $I$ is the sharp image and $N$ is the additive noise. In blind deblurring we are given $B$ only, and our goal is to predict a latent image $L$ which is the closest approximation to the sharp image $I$. This is an ill-posed problem, as we have to predict both $L$ and $K$. Predicting the kernel accurately is essential, else it may lead to various artifacts \cite{odena2016deconvolution}, using learning based approach gives an accurate estimate of blur kernel compared to statistical approaches or skips the kernel estimation process altogether (i.e end-to-end). After estimation of blur kernel the problem converts to non-blind deconvolution, which can be solved using methods(\cite{Schmidt_2013_CVPR}, \cite{Schuler_2013_CVPR})

Statistical methods have another limitation i.e their inability to parallelize because a majority of them rely on coarse to fine iterative methods. Although deep learning models are significantly harder to train but once trained their inference time is comparatively fast. Moreover, deep learning methods have shown better on benchmarking metrics (PSNR and SSIM).   

In this paper we have divided the deep learning methods into two broad categories 
\begin{enumerate}
	\item Estimation of Kernel - Here the proposed deep learning architectures are used to estimate some features (Fourier coefficients\cite{Chakrabarti}, motion flow \cite{Gong_2017_CVPR }\cite{Sun_2015_CVPR}) of the blur kernel or deriving the deconvolution filter \cite{Chakrabarti} which can be used to get back the sharp image.
	
	\item End to End - These methods are kernel free, that means we don’t estimate the blur kernel, rather only the blurred image is required and the model generates the predicted restored image. Some of these methods rely on generative models (\cite{Kupyn_2018_CVPR}, \cite{Ramakrishnan}, \cite{Nah_2017_CVPR}) which are trained in an adversarial method.   
\end{enumerate}

The emphasis of this paper is on the “architecture” proposed by several author instead of the specific details of the architecture and to foster further research in blind deblurring using learning based methods. 

\section{Methods}

\subsection{\textbf{Estimation of Kernel and its Attributes}}

\subsubsection{\textbf{Extraction of Features}}
\begin{figure}[h]
	\centering
	\includegraphics[width=0.5\textwidth]{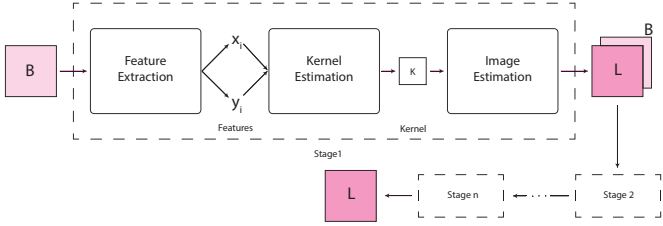}
	\caption{Shows the multi stage architecture used by \textit{Schuler et. al.}\cite{Schuler}, here the different modules in a stage are shown for the first stage only. The latter stages are identical to the first, except the input which is concatenation of blurred image and the restored image of the previous layer}
	\label{fig:sch}
\end{figure}

For a optimal deblurring we require global information i.e data from different parts of the image, but to do so we need to have connectivity with all the pixels of image which will lead to a huge parameter space making it difficult to train and converge, hence \textit{Schuler et. al.}\cite{Schuler} uses CNNs to extract features locally and then combine them to estimate the kernel. For this they use a multi scale (for different kernel sizes), multi stage architecture, where each stage consists of three modules feature extraction, kernel estimation, latent image estimation (Fig. \ref{fig:sch}).  In the first stage given a blurry image, a sharp image is estimated, for later stages they gave the blurry image concatenated with the estimated sharp image of the previous stage as input.

In feature extraction module they used a convolutional layer to extract features using filters $f_j$, then they used $\tanh$ to introduce non-linearity and finally these hidden features are linearly recombined using cofficients $\alpha_{ij}$ and $\beta_{ij}$ to form hidden images $x_i$ and $y_i$ for stage $i$ used for kernel estimation, formally,
\begin{equation}
\begin{split}
	x_i = \sum_{j}\alpha_{ij}\tanh(f_j * y)\\
	y_i = \sum_{j}\beta_{ij}\tanh(f_j * y)
\end{split}
\end{equation}
where $y$ is blurred image $B$ for first stage or concatenation of $B$ and predicted sharp image $L$ for later stages.

Given $x_i$ and $y_i$ the kernel estimation module estimates the kernel $K$ by minimizing,
\begin{equation}
	\sum_{i}\lVert K * x_i - y_i\rVert^2 + \beta_k\lVert K\rVert^2
	\label{eqn:kestm}
\end{equation}
for $K$. Given $K$ we can find the latent (restored) image $L$ by solving the equation,
\begin{equation}
	\lVert K * L - B\rVert^2 + \beta_x\lVert L\rVert^2
	\label{eqn:lestm}
\end{equation}
for $L$, where both $\beta_k$ and $\beta_x$ are regularization weights. Both (Eqn.\ref{eqn:kestm}) and (Eqn.\ref{eqn:lestm}) can be solved in one step in Fourier space.

\subsubsection{\textbf{Estimation of Fourier Coefficients}}
Given a blurry image $B[n]$ where $n \in \mathbb{Z}^2$ are the indexes of pixels. We need to find a latent sharp image $L[n]$ such that it resembles the sharp image $I[n]$ closely where,
\begin{equation}
	 B[n] = (I * K)[n] + N[n]
\end{equation}
where $K[n]$ is the blur kernel such that $ K[n] \geq 0$ (positivity constraint), $\sum_{n}K[n] = 1$ (unit sum constraint) and $N[n]$ the noise.

\begin{figure}[h]
	\centering
	\includegraphics[width=0.5\textwidth]{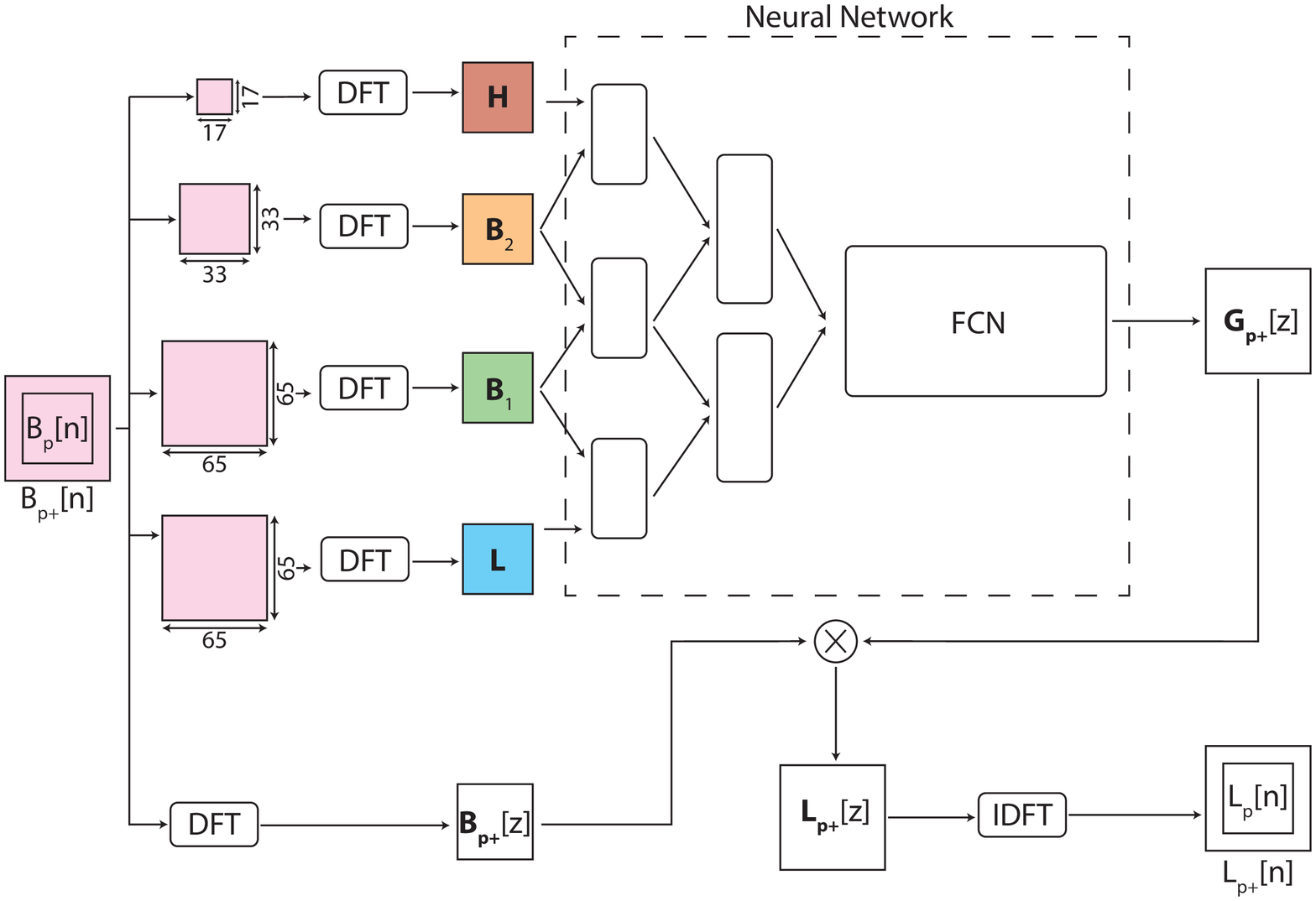}
	\caption{Architecture used by \textit{Chakrabati}\cite{Chakrabarti} for prediction of Fourier coefficients for the deconvolution filter. Here $H$ is high pass, $B_2, B_1$ are band pass, while $L$ is low pass frequency band. The letters in bold are Fourier transforms of the corresponding symbols.}
	\label{fig:chakrabati}
\end{figure}

In the method given by \textit{Chakrabarti} \cite{Chakrabarti} a blurry image $B[n]$ is divided into several overlapping patches. Given a blurry patch $B_p = \{B[n] : n \in p\}$ they considered the surrounding pixels of the patch while finding its Fourier coefficients for better results, let the blurry image with the neighboring pixels be $B_{p^+} = \{B[n] : n \in p^+\}$ where $p \subset p^+$.

Then they used a neural network (Fig. \ref{fig:chakrabati}) to predict the Complex Fourier Coefficients of the Deconvolution Filter $\textbf{G}_{p^+}[z]$ for the blurry patch $B_{p^+}$, where $z$ is the two dimensional spatial frequencies in DFT (Discrete Fourier Transform). Then the filter is applied to the DFT of $B_{p^+}$ i.e $\textbf{B}_{p^+}[z]$ giving us the DFT of latent sharp image $\textbf{L}_{p^+}[z]$,
\begin{equation}
	\textbf{L}_{p^+}[z] = \textbf{B}_{p^+}[z] \times \textbf{G}_{p^+}[z]
\end{equation}
After getting $\textbf{L}_{p^+}[z]$, we can use a inverse discrete Fourier transform (IDFT) to get the latent image patch $L_{p^+}$ from which we can extract $L_p$.

To generate coefficients of the filter they used the architecture shown in (Fig. \ref{fig:chakrabati}).
The architecture uses multi-resolution decomposition strategy i.e the initial layers of the neural network are connected to only adjacent bands of frequency and not fully connected (here they are considering locality in the frequency domain, in contrast to CNNs which consider locality in the spatial domain). The image is sampled into patches of various resolution and a lower resolution patch is used to sample a higher frequency band using DFT. The loss function for the network is,
\begin{equation}
	L = \dfrac{1}{|p|}\sum_{n \in p}(L_p[n] - I_p[n])^2
\end{equation}
They combined all the restored patches to get the first estimate of the latent image $L_N[n]$. It is assumed that the entire image is blurred by the same motion kernel(uniform blur), but they predicted different motion kernels for different patches, hence to find a global motion kernel $K_\lambda[n]$ they used the first estimate  $L_N[n]$ as follows,
\begin{equation}
	K_\lambda = arg\text{ min}\sum_{i}\lVert (K * (f_i * L_N)) - (f_i * B)\rVert^2 + \lambda\sum_{n}|K[n]|
\end{equation}
Here $f_i$ are different derivative filters. They use $L1$ regularization.
In classical statistical methods refining latent image from a previous estimate is an iterative step, while here they only do it once to estimate the global blur kernel. After estimation of the global blur kernel, the problem becomes that of a non-blind deblurring and latent sharp image can be estimated using deconvolution.

\subsubsection{\textbf{Estimation of Motion Vector for each Patch}}
\begin{figure}[h]
	\centering
	\includegraphics[width=0.5\textwidth]{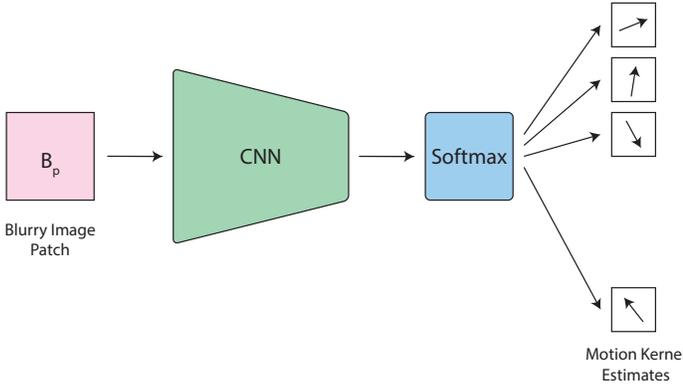}
	\caption{Network architecture for predicting the motion kernel of a given blurred patch used by \textit{Sun et. al.}\cite{Sun_2015_CVPR}}
	\label{fig:sun}
\end{figure}
In this method proposed by \textit{Sun et. al.}\cite{Sun_2015_CVPR} an image is divided into several overlapping patches. For each patch a CNN with a fully connected layer and softmax layer is used to find the probability distribution of motion kernels for that patch (Fig.\ref{fig:sun}). Given a patch $\Psi_{p}$ centered at pixel p, the network finds a probability distribution,
\begin{equation}
	P(m = (l, o) | \Psi_{p})
\end{equation}
where $m = (l, o)$ is the motion kernel with length $l$ and orientation $o$. Here $l \in S^l$ and $o \in S^o$ both $S^l$ and $S^o$ are discretized sets of length and orientation. Due to discretization the number of motion kernels is limited which leads to blocky artifacts. Hence, they rotated the image and its corresponding motion kernel by the same amount to get new data entry, which is then used in training this increases the range of $S^o$ that is given a patch $\Psi_{p}(I)$ of image $I$ and its corresponding motion kernel $m = (l, o)$, if image is rotated by an angel of $\theta$ then for patch $\Psi_{p}(I_{\theta})$ they got the motion kernel as $m = (l, o - \theta)$. Since they are doing a multicalss classification(where each class is a motion kernel) they use cross entropy loss given as,
\begin{equation}
	P(m=(l, o)|\Psi) = \dfrac{\exp(z_i)}{\sum_{k = 1}^{n}\exp(z_k)}
\end{equation}
where $z$ is the output of the last fully connected layer and $n = |S^l|\times|S^o|$ i.e $n$ is the total number of motion kernels.
Since the patches are overlapping many patches may contain the same pixel, in such case the confidence of motion kernel $m$ at a pixel $p$ is given as
\begin{equation}
C(m_p=(l, o)) = \dfrac{1}{Z}\sum_{q:p\in\Psi_{q}}G_{\sigma}(\lVert x_p - x_q \rVert^2)P(m=(l, o)|\Psi_{q})
\end{equation}
where $q$ is center pixel of patch $\Psi_{q}$ such that $p \in \Psi_{q}$. $x_p$ and $x_q$ are the coordinates of $p$ and $q$ respectively. $G_{\sigma}$ is a Gaussian function that gives more weight to patches whose center pixel $q$ is closest to $p$. $Z$ is a normalization constant.

\begin{figure}[h]
	\centering
	\includegraphics[width=0.5\textwidth]{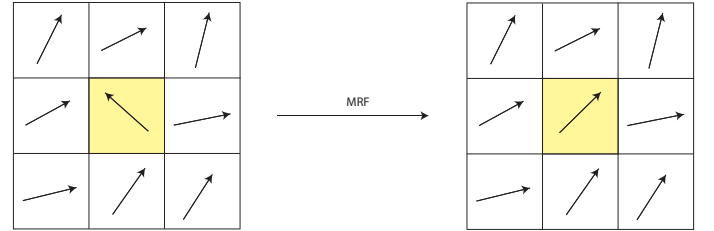}
	\caption{This shows given a pixel $p$ (in yellow) how MRF smoothen its value based on $N(p)$ i.e its neighboring pixels}
	\label{fig:mrf}
\end{figure}

After estimating the motion kernels for all the patches they are concatenated and a Markov Random Function (MRF) is used to merge them all together, smoothen the transition of motion kernels amongst neighboring pixels (Fig.\ref{fig:mrf}) and generates a dense motion field by minimizing energy function,
\begin{equation}
\sum_{p\in\Omega}[-C(m_p=(l_p, o_p))+\sum_{q\in N(p)}\lambda[(u_p-u_q)^2 + (v_p - v_q)^2]]
\end{equation}
where $\Omega$ is a image region. $u_p, u_q, v_p, v_q$ are defined as $u_i = l_i\cos(o_i)$ and $v_i = l_i\sin(o_i)$ for $i = p, q$. $N(p)$ is the neighborhood of $p$. The first term gives more weight to using the motion kernel that the CNN chose with the highest confidence, while the second term looks at the neighboring pixels and tries to smoothen it. After predicting the motion field they deconvolve the blurred image with it to get the deblurred image.

\subsubsection{\textbf{Estimation of Dense Motion Flow for entire Image}}
In the previous approach of \textit{Sun et. al}\cite{Sun_2015_CVPR} a motion kernel was predicted for each patch using a CNN classifier and then all the motion kernels were smoothened using Markov Random Field (MRF) to get the dense motion field. In the method used by \textit{Gong et. al.}\cite{Gong_2017_CVPR} they also predict a dense motion field, but a pixel wise dense motion field is generated for the entire image directly (i.e image is not divided into patches).

\begin{figure}[h]
	\centering
	\includegraphics[width=0.5\textwidth]{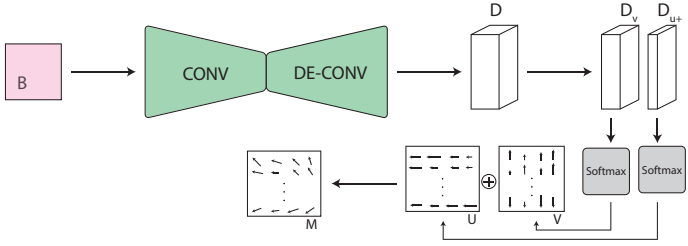}
	\caption{Architecture used by \text{Gong et. al.}\cite{Gong_2017_CVPR} to predict the motion field given a blurry image}
	\label{fig:mbmf}
\end{figure}

In \textit{Sun et. al}\cite{Sun_2015_CVPR} they assumeded uniform motion blur within a single patch as only one motion kernel was chosen for a patch. This does not generalize to real life data properly were we can have a heterogeneous motion blur i.e  motion may vary from pixel to pixel. In such cases an end to end approach of generating motion field \cite{Gong_2017_CVPR} can give better results as they are considering the entire image (larger spatial context) instead of a single patch. Hence, this method is suitable for heterogeneous motion blurs. It does not require any post processing like MRF.

If the network is represented by a function of $f$. Then given a blurred image $B$, the goal of the network is to generate the motion field $M$, i.e
\begin{equation}
f(B) = M 
\end{equation}
where the motion field $M$ can be represented as,
\begin{equation}
M = (U, V)
\end{equation}
where $U$ and $V$ are the horizontal and vertical motion maps respectively.
Now given a pixel $p = (i, j)$ where $(i, j)$ are the coordinates of pixel, then we get,
\begin{equation}
M(i, j) = (U(i, j), V(i, j))
\end{equation}
let $M(i, j) = m_p$, $U(i, j) = u_p$ and $V(i, j) = v_p$ then we get,
\begin{equation}
m_p = (u_p, v_p)
\end{equation}
where $u_p \in \mathbb{D}_u$ and $v_p \in \mathbb{D}_v$. Here $\mathbb{D}_u$ and $\mathbb{D}_v$ are the discretized motion vectors in the horizontal and vertical directions respectively, they are defined as, $\mathbb{D}_u = \{u|u\in\mathbb{Z}, |u| \leq u_{max}\}$ and $\mathbb{D}_v = \{v|v\in\mathbb{Z}, |v| \leq v_{max}\}$.

But, two motion vectors of opposite directions and same magnitude would generate the same blur pattern i.e $m_p = (u_p, v_p)$ and $-m_p = (-u_p, -v_p)$ would give the same blur, hence they restrict the horizontal motion vector to be positive only i.e $u_p \in \mathbb{D}_u^+$ where $\mathbb{D}_u^+ = \{u|u\in\mathbb{Z}^+, |u| \leq u_{max}\}$, this is done by letting $(u_p, v_p) = \phi(u_p, v_p)$ where,
\begin{equation}
\phi(u_p, v_p) = \begin{cases}
(-u_p, -v_p) \text{ if } u_p < 0\\
(u_p, v_p) \text{ if } u_p \geq 0
\end{cases}
\end{equation}
If an image of size $P \times Q$ is sent trough the network(excluding the softmax) it generates a feature map of size $P\times Q\times D$ where $D = |\mathbb{D}_u^+| + |\mathbb{D}_v|$. This feature map is then divided into two parts of shape $P\times Q \times|\mathbb{D}_u^+|$ and $P \times Q \times |\mathbb{D}_v|$. These two features pass through separate softmax layers to generate the horizontal and vertical motion maps $U$ and $V$ receptively. Using these two vector maps they generate the final motion field $M$. After getting the motion field $M$ this becomes a non-blind debluring problem and a deconvolution is used to get the sharp image.

\subsection{\textbf{End to End}}

\subsubsection{\textbf{Without Adversarial Loss}}
Deblurring requires a large receptive field (global knowledge), but CNNs provide local knowledge and do not show the long range dependencies properly, for this reason \textit{Nah et. al}\cite{Nah_2017_CVPR} (refer \ref{adv_loss}) uses a scaled structure and a large number of convolutional layers with residual connections to improve the receptive field of the structure, but this also makes it harder to converge due to the large number of parameters. Hence (\textit{Tao et. al.}\cite{Tao_2018_CVPR}) use a scale recurrent structure where they still use a scaled network, but significantly reduce the number of parameters by using a smaller encoder-decoder type network with a recurrent module and also share the weights between scales.
\begin{figure}[h]
	\centering
	\includegraphics[width=0.5\textwidth]{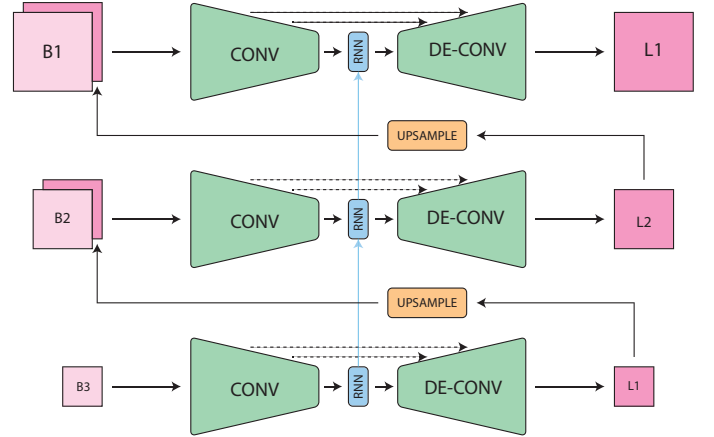}
	\caption{Scale Recurrent Network Architecture used by (\textit{Tao et. al.}\cite{Tao_2018_CVPR})}
	\label{fig:srn}
\end{figure}

Scale recurrent network (Fig.\ref{fig:srn}) \cite{Tao_2018_CVPR} consists of three parts, encoder ($Net_E$), recurrent ($Net_R$) and decoder ($Net_D$) modules. This can be represented as,
\begin{equation}
	\begin{split}
		f^i = Net_E(B^i, L^{i - 1\uparrow}; \theta_E)\\
		h^i, g^i = Net_R(h^{i - 1\uparrow}, f^i; \theta_R)\\
		L^i = Net_D(g^i; \theta_D)\\
	\end{split}
	\label{srn}
\end{equation}
$\theta_E, \theta_R, \theta_D$ are the weights of their respective modules.

The encoder module consists of convolutional layers with residual connections. For the first scale, only the blurred image is used as input, for all the subsequent layers both the blurred image $B^i$ and the restored image of the previous scale $L^{i - 1\uparrow}$ are concatenated and both are sent as input. The encoder module is used to extract features $f^i$, it gradually decreases the length and breadth but increases the number of channels.

The recurrent module can be a vanilla RNN, GRU or LSTM, in \textit{Tao et. al} they used convolutional LSTM (ConvLSTM) which gave the best results. They also trained a network without any Recurrent module and it gave lower performance compared to the one which included a recurrent module. It takes as input the hidden features of the previous scale's recurrent module $h^{i - 1\uparrow}$ and the features generated by the current scales encoder $f^i$. The hidden features of the previous scale passes the intermediate results and blur patterns of the previous scale which benefits the current scale. Gradient clipping is used for this module  only. It gives as output a modified set of features $g^i$ and the hidden features of the current scale $h^i$.

The decoder module consists of a few convolutional layers with residual connections (same dimensions are maintained using padding) followed by a deconvolutional layer which increases the spatial dimensions and decreases the number of channels until the we get latent image for the scale $L^i$.
$\uparrow$ operator (Eqn. \ref{srn}) is used to adapt the dimensions of features or images to that of the next scale. $\uparrow$ can be deconvolution, sub-pixel convolution, image resizing, bilinear interpolation, etc.  

Combining all the three modules a single scale in the network can be represented as, 
\begin{equation}
 L^i, h^i = Net_{SR}(B^i, L^{i -1\uparrow}, h^{i-1\uparrow}; \theta_{SR})
\end{equation}
where $\theta_{SR}$ is the weight shared across all scales.

Scaled Recurrent Network uses Euclidean Loss given below,
\begin{equation}
	 L = \sum_{i = 1}^{n}\dfrac{\kappa_i}{N_i}\lVert L^i - I^i \rVert^2_2
\end{equation} 
where $L^i$ is the latent restored image, $I^i$ is the ground truth sharp image. $\{\kappa_i\}$ are weights for each scale, and $N_i$ is the number of elements in $L^i$ to be normalized.

\textit{Noorozi et. al.}\cite{Noroozi} also uses a three pyramid stages chained together, each consisting of several convolutional and deconvolutional layers $(N_1, N_2, \text{ and } N_3)$ (Eqn. \ref{noorozi}) which recreates a multiscale pyramid schemes which were previously used in the classical methods. The key idea of this pyramid structure is that the downsampled version of the blurred image has less amount of blur and it is easier for removal. Hence the goal of each respective network (or stage) is to mitigate the blur effect at that corresponding scale. It also helps break down the complex problem of deblurring into smaller units.  

\begin{figure}[h]
	\centering
	\includegraphics[width=0.5\textwidth]{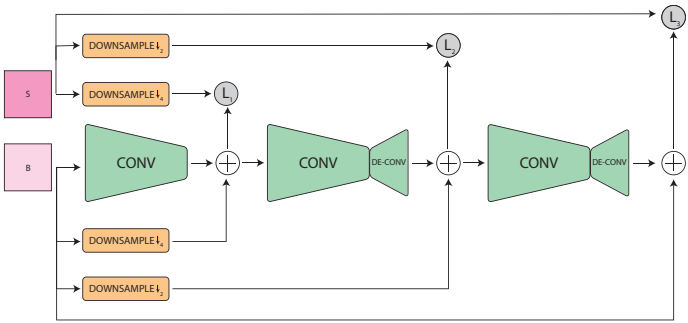}
	\caption{Architecture used by \textit{Noorozi et. al.} \cite{Noroozi}. Here the three CNNs starting from the left denotes $N_1, N_2, N_3$ respectively.}
	\label{fig:noorozi}
\end{figure}

Firstly the blurred image is given as input to the first network $N_1$ (pure convolution) without any downsampling and it’s output is added with the downsampled version of the same blurred image by a factor of four. After this the first loss $L_1$ is calculated using \ref{noorozi} by calculating the difference (or MSE) between the downsampled sharp image and the added sum of the network $N_1$’s output and downsampled blurred image. This same process is repeated for network $N_2$ and $N_3$ for calculating losses $L_2$ and $L_3$ but the downsampling factor are two and one (no downsampling) respectively. The these three computed losses are added resulting in final loss function for this model.

\begin{equation}
	\begin{split}
		L_1 = \sum_{(B, I)} |N_1(B) + d_{1/4}(B) - d_{1/4}(I)|^2\\
		L_2 = \sum_{(B, I)} |N_2(N_1(B) + d_{1/4}(B)) + d_{1/2}(B) - d_{1/2}(I)|^2\\
		L_3 = \sum_{(B, I)} |N_3(N_2(N_1(B) + d_{1/4}(B)) + d_{1/2}(B)) + B - I|^2
	\end{split}
	\label{noorozi}
\end{equation}

The problem in this architecture is when there are extreme blurs,the network leaves the images untouched but it does not suffer from artifacts.

Spatially variant blurs in dynamic scenes require a large receptive field, while CNN have a local knowledge (small receptive field) and spatially invariant weights, to remove this problem they have to use larger networks with more convolutional layer, leading to more parameters which is difficult to train. Hence, the challenge is to have a small architecture with a large receptive field, to this end \textit{Zhang et. al}\cite{Zhang_2018_CVPR} proposes the use of Recurrent Neural Network as a deconvolutional operator which increases the receptive field (long range dependencies).

\begin{figure}[h]
	\centering
	\includegraphics[width=0.5\textwidth]{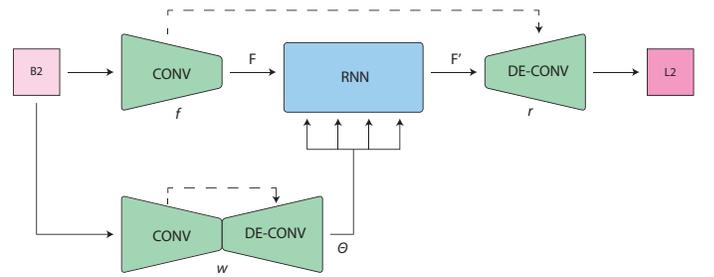}
	\caption{Architecture used by \textit{Zhang et. al.} \cite{Zhang_2018_CVPR}}
	\label{fig:zhang}
\end{figure}

The network proposed by \textit{Zhang et. al.}\cite{Zhang_2018_CVPR} uses three CNN and one RNN. The CNN are used for feature extraction, image reconstruction, and pixel-wise weight generation (for the RNN). While the RNN is used as a deconvolutional operator with a large receptive field (Fig. \ref{fig:zhang}). 
Weight generation for the RNN is done by passing an image through a encoder-decoder architecture CNN. For both the decoder part of the weight generation module and the image restoration module they use bilinear interpolation (no deconvolution) to avoid checkerboard artifact \cite{odena2016deconvolution} The RNN generates receptive fields in one direction (single dimension), hence they use a convolutional layer after every RNN to fuse the receptive fields together to get a two dimensional structure. Skip connections are added to avoid vanishing gradient problem and for faster training.

If $f$ is the feature extraction module, $rnn$ is the RNN module, $w$ is the weight generation module, and $r$ is the restoration module, then the network proposed by \textit{Zhang et. al.}\cite{Zhang_2018_CVPR} can be summarized as,
\begin{equation}
	\begin{split}
		F = f(B)\\
		\theta = w(B)\\
		F^{\prime} = rnn(F; \theta)\\
		L = r(F^{\prime})
	\end{split}
\end{equation}
where $B$ is the blurry image, $F$ is the extracted features, $\theta$ is the pixel-wise generated weights, $F^{\prime}$ are the modified features after passing through the RNN, and $L$ is the latent (predicted) deblurred image. 

\subsubsection{\textbf{With Adversarial Loss}} \label{adv_loss}

\begin{figure}[h]
	\centering
	\includegraphics[width=0.5\textwidth]{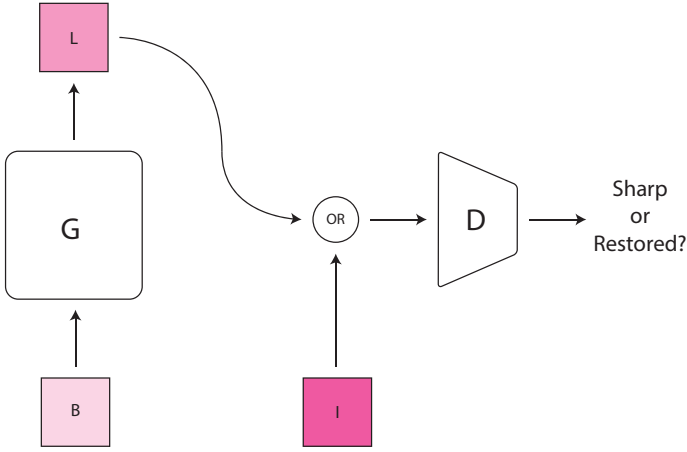}
	\caption{The basic structure of a GAN, where $G$ denotes the Generator and $D$ denotes the discriminator.}
	\label{fig:gan}
\end{figure}

Blind Deblurring can also be solved end-to-end by generative models like Generative Adversarial Networks \cite{Goodfellow}\cite{Isola_2017_CVPR}\cite{Arjovsky}. The approach Generative Adversarial Networks take is to have two different agents play a game against each other. One of the agents is a generator network which tries to generate data and the other is a discriminator network which examines data and checks whether it came from the real distribution (ground truth sharp image) or model distribution (restored blurred image). The goal of the generator is to fool the discriminator into believing that its output is from the real distribution. These generator and discriminator modules are neural networks whose parameters can be tuned by backpropagation and as both players get better at their job over time eventually the generator is forced to create data as realistic as possible.

\begin{figure}[h]
	\centering
	\includegraphics[width=0.5\textwidth]{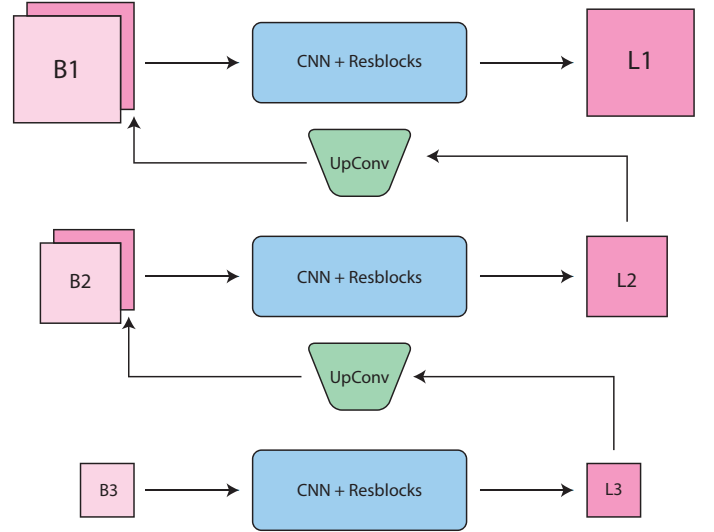}
	\caption{Multiscale architecture used by \textit{Nah. et. al} \cite{Nah_2017_CVPR}}
	\label{fig:mul}
\end{figure}

\textit{Nah et. al.}\cite{Nah_2017_CVPR} also uses a Multiscale Convolutional Neural Network i.e coarse(low resolution) to fine(high resolution) structure. The blurred and sharp images are scaled down to form a Gaussian pyramid structure, this is done because convolution can only capture local information, hence lower resolution images are used to capture the long range dependencies, whereas the high resolution images are used to capture the fine grained details. Each of these scaled blurred images passes through a layer of multiple convolutional and residual blocks (residual blocks enable training in large networks without over fitting) to generate the corresponding latent image for that scale, then for each scale MSE(Mean Squared Error) with the sharp image is calculated and back propagation is done. The MSE for all the scales are averaged together to give the content loss as follows:
\begin{equation}
	L_{content} = \dfrac{1}{2K}\sum_{k=1}^{K}\dfrac{1}{c_kh_kw_k}\lVert L_k - I_k\rVert^2 
\end{equation}
Here $K$ is the total number of scales, $c_k, h_k, w_k$ are the channels, height and width of the $k^{th}$ scale, and $L_k$ and $I_k$ are the latent and sharp images of the $k^{th}$ scale respectively.

The output of current scale is given as input to the next scale. The next scale is of a higher resolution, hence the latent image of the current scale is passed through a upconvolutional(transpose convolution) layer and is concatenated with the blurred image input of the next layer. Except for the last layer whose output latent image is the same size as the original image, hence does not need any upconvolution. This generated deblurred image of the last scale is given as input to a discriminator or some sharp image is given as input, and the discriminator tells weather the image given was originally sharp or was deblurred by the Multi-Scaled Network. Discriminator loss function (Adversarial Loss) \cite{Goodfellow} is as follows:
\begin{equation}
	L_{adv} = \mathbb{E}_{S\sim p_{sharp}}[\log(D(S))] + \mathbb{E}_{B \sim p_{blurred}}[1 - \log(G(B))]
\end{equation}
Here $D$ is the Discriminator i.e a CNN classifier and $G$ is the Generator i.e our Multi-Scaled CNN. Generator $G$ for each scale can be defined as,
\begin{equation}
L^i = G(B^i, L^{i - 1\uparrow}; \theta_i)
\end{equation}
where $L^i, B^i$ is the generated and blurred image for $i^{th}$ scale respectively. $L^{i - 1\uparrow}$ is the generated image of the previous scale where $\uparrow$ denotes the upconvolution function used to upscale the dimension of $L^{i -1}$ to be same as $L^i$. $\theta_i$ are the weights of $i^{th}$ scale.

Combing both loss functions i.e Content and Adversarial Loss we get:
\begin{equation}
	L_{total} = L_{content} + \lambda L_{adv}
\end{equation}
Where $\lambda$ is a weight constant. 

\textit{Ramamkrishnan et. al}\cite{Ramakrishnan} also uses an adversarial way of training, but the generator uses a structure similar to \textit{DenseNet}\cite{Huang_2017_CVPR} with a global skip connection.
Similar to \textit{Nah et. al.}\cite{Nah_2017_CVPR}, here dimensions are maintained throughout the convolutional layers, so that no deconvolution module needs to be used, preventing checkerboard effect \cite{odena2016deconvolution}.
Using a densely connected CNN in generator reduces the vanishing gradient problem, strengthens feature propagation and reuse and reduces the number of parameters, all of which in turn allows us to use a smaller network with smoother training and faster inference time.

The generator is divided into three parts, head, dense field and tail.
Head creates sufficient activations for the dense field using convolutional layer.
The dense field consists of several dense blocks, each dense block has a ReLU to add non linearity, a 1$\times$1 convolution to limit the number of activaitons (or channels), a convolutional layer (3 $\times$ 3), and batch normalizations. In \textit{DenseNet}\cite{Huang_2017_CVPR} $l^{th}$ layer of convolution is connected to the features of all the previous layers as opposed to the immediately previous layer (like in classic CNN). This dense connectivity is achieved in the generator by concatenating the output of $l^{th}$ layer with the output of $(l + 1)^{th}$ layer, which in turn concatenates its output with the output of next layer i.e $(l +  2)^{th}$ layer and so on.
All the convolutional layers in the dense filed use spatial and dilated convolution alternatively, this increases the receptive field of the network while still restraining the number of parameters to be learned. The rate of dilation increases till the middle layer and then decreases till the tail is reached.
Tail adds non-linearity and uses $1\times1$ convolution to reduce the number of activations.

They added the output of head to the output of tail to form a global skip connection, this allows gradients to flow back to the first layers of convolution which helps in learning (gradient updates) of the lower layers. Shorter connections between layers close to the output and layers close to the input, results in better accuracy and efficiency.

Similar to \textit{Nah et. al.}\cite{Nah_2017_CVPR} here the loss functions is also divided into two parts but with slight differences i.e instead of finding the MSE between predicted and sharp image (content loss), they find the MSE between the features (taken from end layers of a pretrained VGG16 network) of the predicted and sharp image. This is known as Perceptual Loss \cite{Johnson2016Perceptual}.
\begin{equation}
	L_{precep} = \dfrac{1}{W \times H}\sum_{x = 1}^{W}\sum_{y = 1}^{H}(\phi(I)_{x, y} - \phi(L)_{x, y})
	\label{precep_loss}
\end{equation}
where $\phi$ denotes the function used to generate the features. $W, H$ are dimensions of the features. $L$ is the predicted latent image i.e $L = G(B)$ for generator $G$ and blurry image $B$.

Instead of simple adversarial loss they use a conditional adversarial loss \cite{Isola_2017_CVPR} i.e with every sharp or predicted image, they also send the corresponding blurred image. Then calculate the probability of weather the image is deblurred or sharp given the blurred image.
\begin{equation}
	L_{adv_{con}} = -\mathbb{E}_{b \in B}[\log D(G(B)|B)]
\end{equation}
where $D$ is the discriminator.

While combining both the losses \textit{Zhang et. al}\cite{Zhang_2018_CVPR} also adds a $L1$ loss which was not present in \textit{Nah et. al}\cite{Nah_2017_CVPR},
\begin{equation}
	L_{total} = L_{precep} + K1\times L_{adv_{con}} + K2\times L_{L1}
	\label{total_loss}
\end{equation}
where $L_{L1}$ is the $L1$ loss. $K1, K2$ are the weight constant.

\text{Kupyn et. al.}\cite{Kupyn_2018_CVPR} uses a method also based on conditional GANs \cite{Isola_2017_CVPR} similar to \textit{Ramakrishnan et. al.}\cite{Ramakrishnan}, the number of layer are significantly less compared to \textit{Nah et. al.}\cite{Nah_2017_CVPR}, decreasing the number of trainable parameters and hence resulting in decrease training time and faster inference time. Instead of  using the conventional loss function of GANs, they used the wasserstein (or called Earth-Mover) distance with gradient penalty \cite{Arjovsky} which has proved to show stability from vanila GANs\cite{Goodfellow} which suffer from mode collapse and  vanishing gradients.
The generator architecture is different from \textit{Ramakrishnan et. al.}\cite{Ramakrishnan} as they use some convlutional layers followed by a series of residual blocks and finally some deconvolution layers.

The generator takes the blurred image as input and produces it’s sharp estimate. The discriminator then tries to model the differences in sharp images (real distribution) and restored image (model distribution) by the generator by computing the Wasserstein distance (Earth mover distance)\cite{Arjovsky}. The perceptual loss is same as in (Eqn. \ref{precep_loss}).
 
The goal is to minimize this entire loss function (which is same as (Eqn. \ref{total_loss}) but without the $L1$ loss) such that the generator is producing well enough restored image from the blurred image and the discriminator network to unable to distinguish the real sharp image (real data distribution) and restored image (model distribution) resulting in output of ½ probability by the discriminator most of the time. This is when the model is said to have converged. 

\section{Performance Evaluation}

\subsection{\textbf{Metric}}
The metrics used to measure similarity between the restored image and the blurred image are Peak Signal to Noise Ratio (PSNR) and Structural Similarity(SSIM). We also compare time taken by different architectures to deblur a blurry image after they are trained (inference time) (Table \ref{table:time}).

PSNR can be thought of as the reciprocal of MSE (Mean Squared Error). MSE can be calculated as,
\begin{equation}
	MSE = \dfrac{\sum_{P, Q}(I - L)^2}{P\times Q}
\end{equation}
where $P, Q$ are the dimensions of the image. $I$ and $L$ are the sharp and deblurred image respectively.
Given MSE, PSNR can be calculated using,
\begin{equation}
	PSNR = \dfrac{m^2}{MSE}
\end{equation}
where $m$ is the maximum possible intensity value, since we are using 8-bit integer to represent a pixel in channel, m = 255.

SSIM helps us to find the structural similarity between two image, it can be calculated using,
\begin{equation}
	SSIM(x, y) = \dfrac{(2\mu_x\mu_y + c_1)(2\sigma_{xy} + c_2)}{(\mu_x^2 + \mu_y^2 + c_1)(\sigma_x^2 + \sigma_y^2 + c_2)}
\end{equation}
where $x, y$ are windows of equal dimension for $B, I$ respectively.
$\mu_x, \mu_y$ denotes mean of $x,y$ respectively. $\sigma_x, \sigma_y$ denotes variance for $x, y$ respectively, whereas $\sigma_{xy}$ is the covariance between $x$ and $y$. $c_1$ and $c_2$ are constants used to stabilize the division.

\subsection{\textbf{K\"{o}hler Dataset}}
K\"{o}hler Dataset \cite{Kohler} consists of 4 images which are blurred using 12 different blur kernels giving us a total of 48 blurred images. To generate the blurred kernels, 6D camera motion is recored and then replayed using a robot, for each image. While replying, the 6D motion is approximated into a 3D motion by considering translation in one plane, and rotation on the plane perpendicular to it. This helps us to approximate actual camera shakes that occur in real life. For more details refer to \cite{Kohler}. The PSNR and SSIM for different deblurring architecture in K\"{o}hler dataset is shown in (Table \ref{table:kohler}).

\begin{table}[h]
	\begin{center}
		\caption{K\"{o}hler Dataset}
		\label{table:kohler}
		\begin{tabular}{|c | c | c |} 
			\hline
			\textbf{Methods}& \textbf{PSNR} & \textbf{SSIM} \\  
			\hline\hline
			\textit{Kupyn et. al.} & 26.10 & 0.816  \\ 
			\hline
			\textit{Tao et. al.} & 26.80 & \textbf{0.838} \\
			\hline
			\textit{Nah et. al.} & 26.48 & 0.812  \\
			\hline
			\textit{Gong et. al.} & 26.59 & 0.742  \\  
			\hline
			\textit{Ramakrishnan et. al.} & \textbf{27.08} & 0.751 \\
			\hline
			\textit{Sun et. al.} & 25.22& 0.774\\
			\hline
		\end{tabular}
	\end{center}
\end{table}

\subsection{\textbf{GoPro Dataset}}
Here, a high resolution (1280 $\times$ 720), high frame rate (240 frames per second) camera (GoPro Hero5 Black) is used to capture video outdoors. To generate blurred image an average of a few frames (odd number picked randomly from 7 to 23) is taken, while the central frame is considered as the corresponding sharp image.
To reduce the magnitude of relative motion across frames they are down sampled and to avoid artifacts caused by averaging we only consider frames were the optical flow is at most 1.
The PSNR and SSIM for different deblurring architecture in GoPro dataset is shown in (Table \ref{table:gopro}).

\begin{table}[h]
\begin{center}
	\caption{GoPro Dataset}
	\label{table:gopro}
	\begin{tabular}{|c | c | c |} 
		\hline
		\textbf{Methods}& \textbf{PSNR} & \textbf{SSIM} \\  
		\hline\hline
		\textit{Kupyn et. al.} & 28.7 & \textbf{0.958}  \\ 
		\hline
		\textit{Zhang et. al.} & 29.2 & 0.931  \\
		\hline
		\textit{Tao et. al.} & \textbf{30.1} & 0.932 \\
		\hline
		\textit{Nah et. al.} & 29.2 & 0.916  \\
		\hline
		\textit{Gong et. al.} & 26.1 & 0.863  \\  
		\hline
		\textit{Noorozi et. al.} & 28.1 & - \\
		\hline
		\textit{Sun et. al.} & 24.6 & 0.842\\
		\hline
	\end{tabular}
\end{center}
\end{table}

\begin{table}[h]
	\begin{center}
		\caption{Inference Time}
		\label{table:time}
		\begin{tabular}{|c | c|} 
			\hline
			\textbf{Methods}& \textbf{Time(sec)} \\  
			\hline\hline
			\textit{Kupyn et. al.} & \textbf{0.8} \\ 
			\hline
			\textit{Tao et. al.} & 1.6 \\
			\hline
			\textit{Nah et. al.} & 4.3  \\
			\hline
			\textit{Zhang et. al.} & 1.4 \\
			\hline
		\end{tabular}
	\end{center}
\end{table}

\section{Conclusion}

We observe that end-to-end methods (\cite{Kupyn_2018_CVPR}, \cite{Nah_2017_CVPR}, \cite{Ramakrishnan}, \cite{Noroozi}, \cite{Tao_2018_CVPR}, \cite{Zhang_2018_CVPR}) have higher PSNR and SSIM compared to methods that estimate the blur kernel (\cite{Gong_2017_CVPR}, \cite{Sun_2015_CVPR}), this is because an error in kernel estimation can lead to various artifacts in image, degrading the restoration.

We also observed that most of the methods tried to increase their receptive field, which allowed long range spatial dependencies, essential for non uniform blurs.

Another motivation was to reduce the size of network and the number of parameters, resulting in faster inference, as it can be clearly seen in (Table \ref{table:time}) that \textit{Nah et. al.}\cite{Nah_2017_CVPR} which has a large network size is slower compared to other networks.

Decreasing the size of network, while maintaining a large receptive field is one of the biggest challenge in learning based deblurring methods.

\bibliographystyle{IEEEtran}
\bibliography{refs}

\end{document}